# IDENTIFYING SEASONAL STARS IN KAURNA ASTRONOMICAL TRADITIONS

Duane W. Hamacher

*Nura Gili Indigenous Programs Unit, University of New South Wales, Sydney, NSW, 2052, Australia*
E-mail: d.hamacher@unsw.edu.au

**Abstract**

Early ethnographers and missionaries recorded Aboriginal languages and oral traditions across Australia. Their general lack of astronomical training resulted in misidentifications, transcription errors, and omissions in these records. Additionally, many of these early records are fragmented. In western Victoria and southeast South Australia, many astronomical traditions were recorded, but curiously, some of the brightest stars in the sky were omitted. Scholars claimed these stars did not feature in Aboriginal traditions. This under-representation continues to be repeated in the literature, but current research shows that some of these stars may in fact feature in Aboriginal traditions and could be seasonal calendar markers. This paper uses established techniques in cultural astronomy to identify seasonal stars in the traditions of the Kaurna Aboriginal people of the Adelaide Plains, South Australia.

## 1  Introduction

In the astronomical traditions of Aboriginal Australians, the rising and setting of particular stars at dusk and dawn are used as calendric markers, noting the changing of seasons, the availability of food sources, and the breeding cycles of animals (Clarke, 2007; Hamacher, 2012; Johnson, 1998; Tindale, 1983). Some calendric stars are described in Aboriginal traditions but their identity is often unclear, in part because of misidentifications (e.g. Howitt, 1884a: 198).

The Adelaide Plains of South Australia are the traditional lands of the Kaurna[1] people (Figure 1). Much of their culture was damaged by colonisation, particularly from 1836 onwards (Manning, 2002). Kaurna astronomical traditions are not well recorded, as the Kaurna "carefully conceal them [astronomical traditions] from Europeans, and even their own males are only at a certain age initiated into the knowledge of them," (Teichelmann, 1841).

What we do know about Kaurna language comes predominantly from the records of the German Lutheran missionaries Christian Gottlieb Teichelmann (1807-1888) and Clamor Wilhelm Schürmann (1815-1893), who came to Adelaide on October 1838. In 1840, they published a dictionary of 2,000 Kaurna words (Teichelmann and Schürmann, 1840). Much of the Kaurna language was provided by local Aboriginal men, including Mullawirraburka (aka King John) and Kadlitpinna (aka Captain Jack), (Amery, 2000: 56).

Teichelmann continued working on the Kaurna language, compiling a manuscript of Kaurna vocabulary and grammatical notes (Teichelmann, 1857). Clarke (1990, 1997) published research on Kaurna astronomical traditions, but the identity of many of the stars remains a mystery.

We do know that in Kaurna traditions, particular stars govern seasons. For example, autumn (*Parnatti*) is signaled by the morning appearance of the star *Parna*. This warns the Kaurna that the annual autumn rains will soon arrive and that they need to build large, waterproof huts (Teichelmann and Schürmann, 1840).

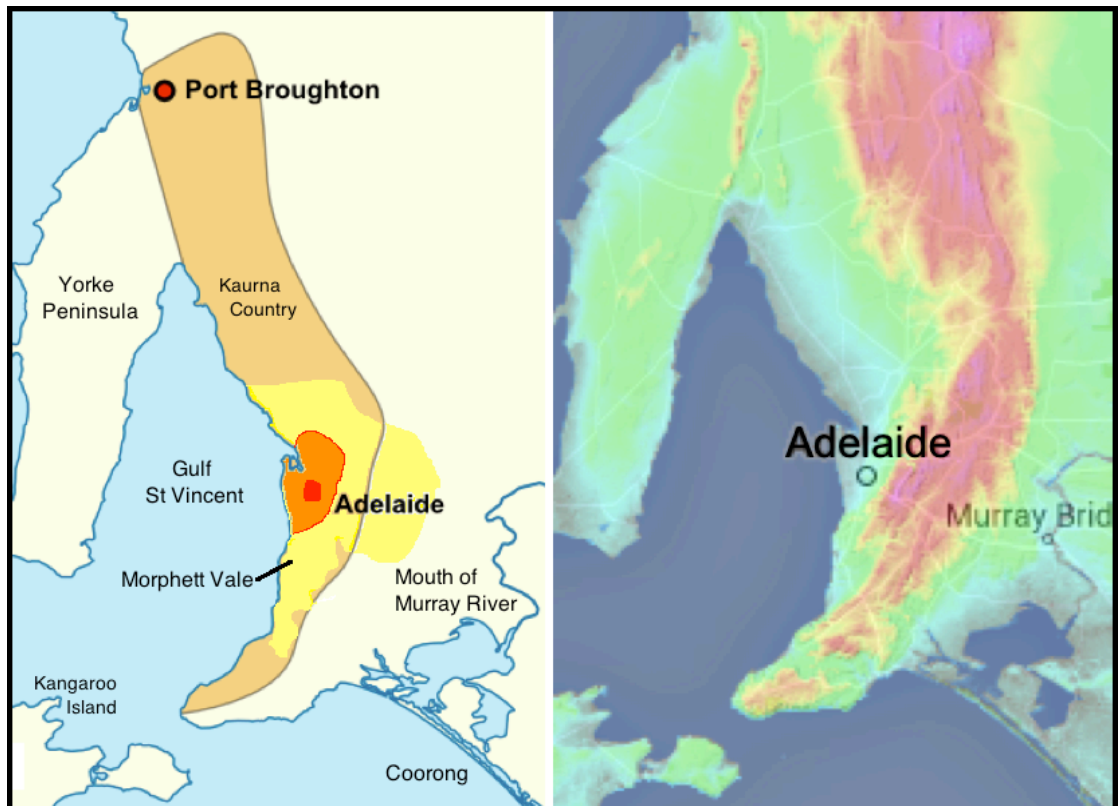

*Figure 1: Left: The traditional lands of the Kaurna people, South Australia, stretching from Adelaide in the south to Port Broughton in the north. Adelaide CBD in red, greater city in orange, and metropolitan area in yellow. Image: Wikipedia Commons license. Right: Topographic map showing the extent of the Adelaide Plains, following the Kaurna lands closely (green indicates low elevation areas while red/pink indicate higher elevation). Image: topographic-map.com.*

Summer (*Woltatti*), the hot season, is governed by *Wolta*, the wild turkey "constellation" [2], and Spring (*Willutti*) is under the influence of *Wilto*, the eagle star. Winter (*Kudlilla*), the rainy season, was not associated with any particular star in the record (Clarke, 1997: 137; Teichelmann and Schürmann, 1840: 4, 12, 55, 57, 58).

## 2      Kaurna Astronomical Traditions

Astronomical traditions of the Adelaide region were first recorded by Teichelmann and Schürmann as part of their linguistic study of the Kaurna language. Since then, a number of publications have explored the local language and astronomy (e.g. Black, 1920; Clarke, 1997, 2007; Gell, 1842; Hartland, 1898; Teichelmann and Schürmann, 1840; Tindale, 1937, 1974, 1983). However, surprisingly little detail about Kaurna astronomy was recorded. The records provide some details (Table 1), but most star names are given with no reference to their Western counterpart, and only fragmentary details are given about the stories of the stars themselves.

We know the Kaurna have rich astronomical knowledge. Gell (1842: 116-117) says that "most of the stars have some legend attached to them" and Teichelmann (1841: 9) wrote that

> *"the exaltation of almost every constellation they give the history of the attending circumstances, which the reasons of their present movements explain."*

The sky was seen as a component of the land; a reflection.

> *"[…] all the celestial bodies were formerly living upon Earth, partly as animals, partly as men, and that they left this lower region to exchange for the higher one. Therefore all the names they apply to the beings on Earth, they apply to the celestial bodies […],"* Teichelmann (1841: 4).

But we also know that the Kaurna were very secretive about their astronomy, only providing fragments of information to the white colonists. Schürmann's Aboriginal informants closely guarded their secrets. When an Aboriginal man told Schürmann about his cosmology, he did so under the condition that he would not tell another Aboriginal person (Schürmann, 1840).

In Kaurna traditions (Gell, 1842: 116; Teichelmann and Schürmann, 1840: 37), *Tinniinyaranna* are the stars of Orion (most likely the belt and scabbard, as with many Aboriginal cultures)[6], representing a group of boys who hunt kangaroo and emu on the celestial plain, while the *Mangkamangkaranna* (the Pleiades star cluster) represents a group of girls digging roots. The red star *Madletaltarni*, probably Betelgeuse (Alpha Orionis) or Aldebaran (Alpha Taurii), is the mother of the Tinniinyaranna and *Parnakkoyerli* is the father, described as "a seasonal star", possibly Rigel. Betelgeuse and Rigel flank the boys on either side (Figure 2).

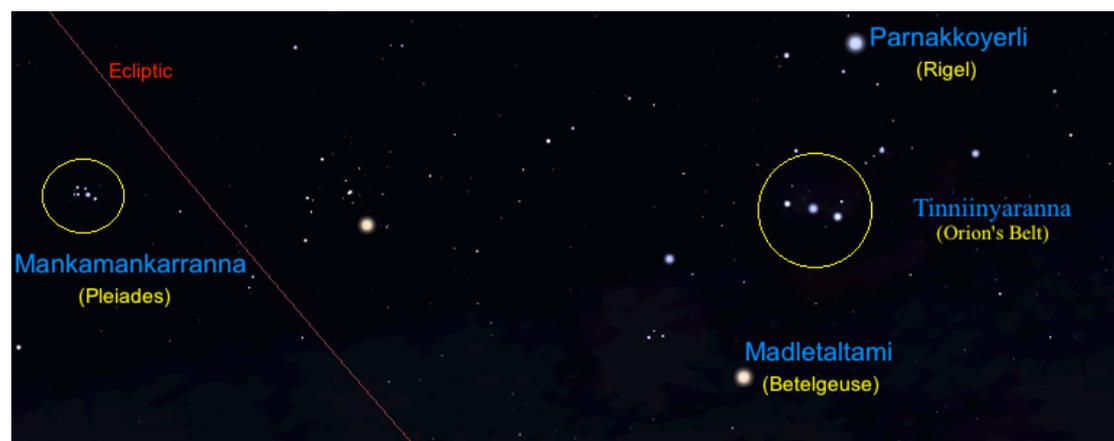

*Figure 2: Stars identified by name in Kaurna astronomical traditions (or inferred by Teichelmann and Schürmann). In Western terminology, Orion is to the right, the Pleiades to the left, and Aldebaran and the Hyades in between.*

Parnakkoyerli means "father of Parna". If Parnakkoyerli is the father of Parna, is he also the father of the Tinniinyaranna, or is this a transcription error? The Western counterparts of Madletaltarni and Parnakkoyerli remain a mystery.

Table 1: Celestial objects in the Kaurna sky. Compiled from Teichelmann and Schürmann (1840) and Clarke (1990, 1997).

| **Kaurna Name** | **Western Name** | **Description** |
| --- | --- | --- |
| Kakirra | Moon | Called Piki in eastern communities. Husband of Tindo (sun), first to ascend into the sky. Encouraged others to follow him to keep him company. |
| Kumomari | Constellation, *Unidentified* | |
| Kurkukurkurra | Orion | Same as Tinniinyaranna, the constellation Orion (probable reference to belt and scabbard stars) |
| Madletaltarni | Red Star, *Unidentified* | Mother of the Tinniinyaranna. Probably Betelgeuse. |
| Mankamankarranna | Pleiades | Girls who dig for roots |
| Mattinyi | Constellation, *Unidentified* | |
| Monana | *Unidentified* | Man who threw spears into sky |
| Ngaiera | Sky | |
| Ngakallamurro | Magellanic Cloud (one of) | Ashes of Parakeets (Adelaide Crimson Rosella?), from *murro* = ashes. The birds are gathered there by another constellation, and then roasted. The constellation is not identified. Teichelmann (1841: 2) suggests both clouds are the lorikeet ashes. |
| Njengari | *Unidentified* | Man who created landscape, then transformed into star. |
| Parna | Star, *Unidentified* | A star indicating Autumn |
| Parnakkoyerli | Star, *Unidentified* | Father of the Tinniinyaranna. Name translates to 'father of Parna.' Probably Rigel. |
| Purle | Star, *Unidentified* | Generic term for a star? |
| Tindo | Sun | Wife of moon. She beats him to death every month but he comes back to life. This describes the lunar cycle. |
| Tinniinyaranna | Orion (Belt/Scabbard?) | Group of young boys who hunt kangaroos, emus, and other game, on the great celestial plain, probably represented by the belt and maybe scabbard of Orion. |
| Wayakka | Star or Constellation, *Unidentified* | |
| Willo | Star, *Unidentified* | One whose older brother (Yunga) has died |
| Wilto | Star, *Unidentified* | Summer season, Eaglehawk (Wedge Tailed Eagle). Possibly same as Willo. |
| Wodliparri | Milky Way | Large River with Reeds |
| Wolta | Star? *Unidentified* | Wild Turkey (Australian Bustard) |
| Womma | Space/Sky | Celestial plain, where the Tinniinyaranna hunted emu and kangaroo. |
| Yurakauwe | Absorption nebulae | Dark spots in the Milky Way, thought to be large ponds in the Wodliparri, and the residence of the aquatic monster Yura. Yura punishes those who break sacred law. |
| Makko | Cloud | |
| *Unnamed* | Planets (generic) | Wives of the sun-father? |
| *Unnamed* | Comets | Evil sisters of the sun? |
| *Unnamed* | Meteors | Orphans |

**3      Seasons on the Adelaide Plains**

The fertile Adelaide Plains of South Australia (SA) are flanked by the Mount Lofty Ranges to the east and Gulf St Vincent to the west, the lower section of which is taken up by the city of Adelaide. Rain is not evenly distributed throughout the year [3] (Figure 3). Long-term data shows that February, the driest and hottest month of the year, only sees an average of 3.6 days and 15.8 mm of rain. June, the wettest month of the year, experiences 14.8 days and 80.6 mm of rain. The rainfall increases in autumn, and decreases in spring.

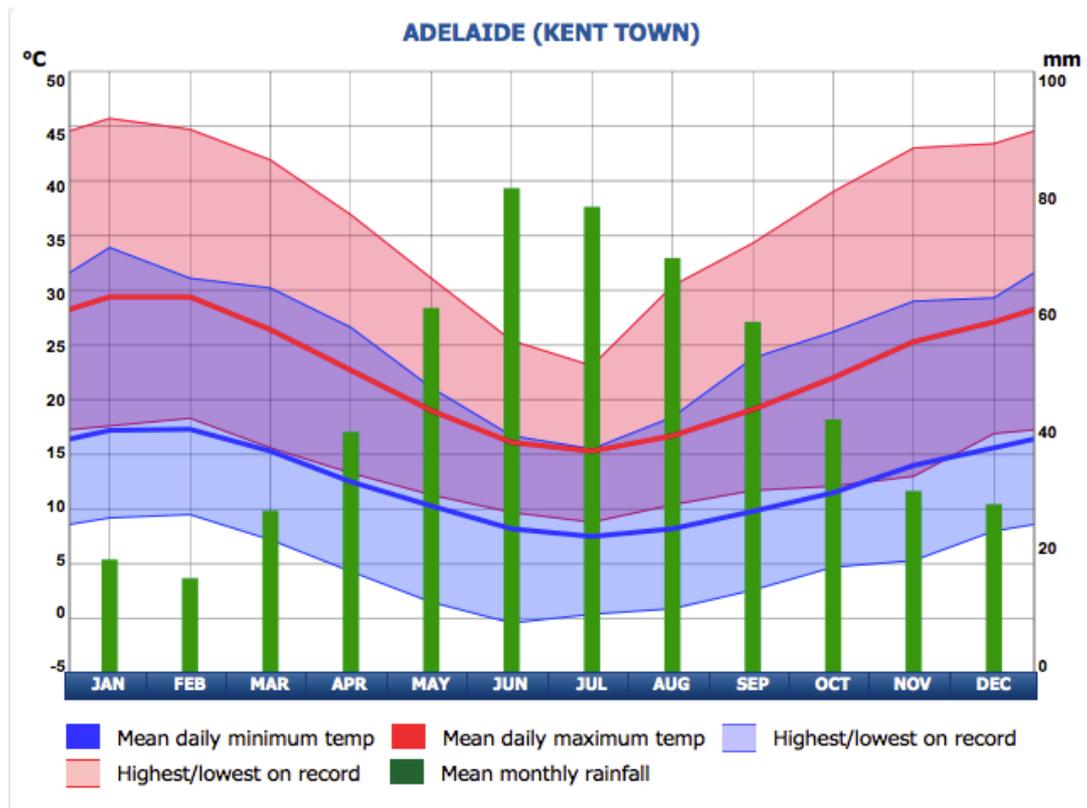

Figure 3: Long-term average seasonal data from Adelaide, recorded between 1977-2014, including information on the temperature, rainfall. From Weatherzone based on data from the Bureau of Meteorology (www.weatherzone.com.au/climate/station.jsp?lt=site&lc=23090)

Ethnographic records indicate four distinct seasons in Kaurna traditions, similar to other Aboriginal groups in southeastern Australia (see Clarke, 1997; Stanbridge, 1861; Teichelmann and Schürmann, 1840). There is ongoing debate as to the reliability of this interpretation, as the traditions and language were recorded by German missionaries and contain their biases and interpretations. For example, Heyes (1999) suggests there were more than four seasons, but due to the backgrounds and research methods of Teichelmann and Schürmann, the Kaurna and European seasons were conflated. This may also be the case with other Aboriginal seasons recorded in southeast Australia that seem to correspond the four European seasons.

Heyes (1999: 16) suggests that the missionaries melded multiple seasons into a single cycle to more closely match the four-season familiar European framework (Figure 4):

- Woltatti (hot season) or Bokarra (hot north winds that blow in summer)

- Wadlworngatti (time of building huts against fallen trees) or Parnatti (Autumn - when the Parna star appears)
- Kudlilla (winter - icy cold south west winds)
- Wullutti (spring – Wilto/Willo star appears).

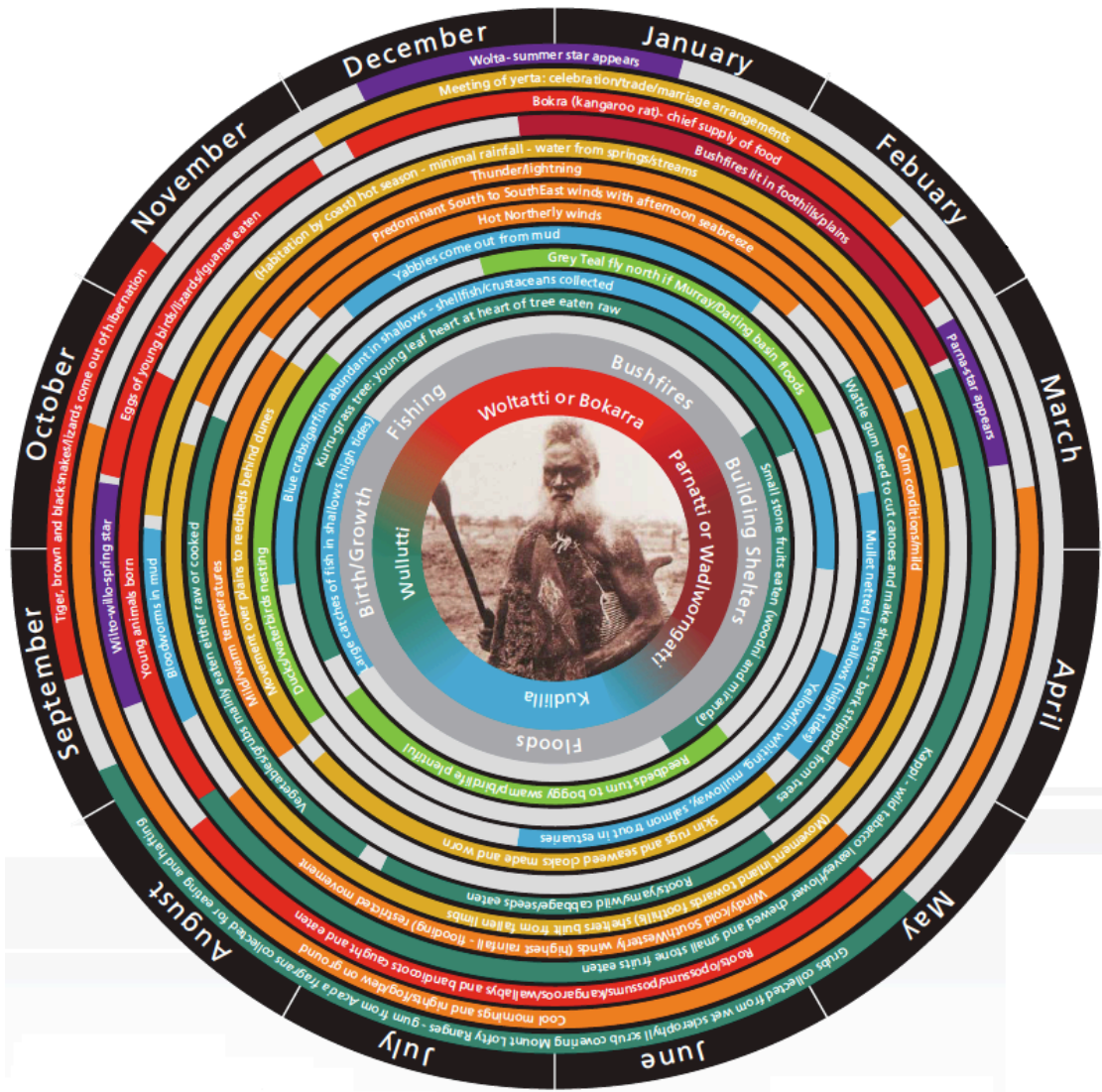

Figure 4: A seasonal Kaurna calendar, based on the Honours research of Scott Heyes (1999). Seasonal stars and the time of their appearance are noted in purple. Image by Scott Heyes and Philip Easson.

Teichelmann and Schürmann attribute summer to *Woltatti*, autumn to *Parnatti*, winter to *Kudlilla*, and spring to *Willutti*.

The stars or constellations associated with Parna (autumn star), Wolta (wild turkey), and Wilto (eagle) are not identified in the literature by their Western counterpart. Clarke (1997) suggests that Kudlilla (winter) is associated with a star, but the original records never explicitly state this. Only Parna is noted as rising at dawn. We are given no information about the rise/set association of the other seasonal stars, so we apply the same criteria to them as we do to Parna: we assume they rise at dawn at the start of their associated season.

In this paper, we apply methods for identifying these stars by examining their heliacal rise time and corresponding seasonal change, and by exploring their presence and/or associations in Kaurna astronomical traditions. The process of identifying previously unnamed celestial objects in Indigenous astronomical traditions can be difficult, so developing and demonstrating rigorous methods for accomplishing this is useful for future work in cultural astronomy.

The published records of Kaurna astronomical traditions are incomplete and may contain errors. Only Parna is clearly denoted as rising at dawn (heliacal rising) to signal the coming winter rains. Neither Wolta nor Wilto are said to rise at dawn to signal seasonal change, and Kudlilla is never specifically associated with a star. No further information is given so it is assumed, for the purposes of this study, that they rise heliacally. By extrapolating the relationship between Parna and seasonal change, we apply criteria to search for candidate stars for Wolta, Wilto, and Kudlilla.

## 4    Methodology

For this study, two important factors are necessary to determine the identity of Kaurna seasonal stars: 1) the heliacal rise time and location of stars, and 2) the time of seasonal change in the Adelaide Plains. The former can be calculated very precisely (to the day), while the latter cannot (ranging from weeks to months). Therefore, a suitable methodology must be developed to identify the best candidates for Kaurna calendric stars.

### 4.1    Calculating Heliacal and Acronycal Rising

When a star first appears on the eastern horizon just before sunrise (before its light is drowned out by the sun), it is referred to as *heliacal rising*. Cultures across the globe have used the heliacal rising of particular stars for hunting/gathering, agriculture, and seasonal markers for millennia (Aveni, 2003; Kelley and Milone, 2011), including Australia (Johnson, 1998). *Acronycal rising* (meaning opposite the sun) is when a star rises in the east at sunset. This is also significant for denoting calendric changes in Aboriginal astronomical traditions (e.g. Hamacher, 2012: 70-86).

The heliacal or acronycal rising of a star is dependent on a number of factors, including the azimuth and altitude of a star, the altitude and azimuth of the sun relative to the star's position in the sky, the location and elevation of the observer, atmospheric conditions, and the star's brightness and colour.

### 4.1.1    Star Brightness

First magnitude stars are the 22 brightest stars in the sky, ranging from the brightest,

*Journal of Astronomical History and Heritage, Vol. 18(1), Preprint*Sirius (Alpha Canis Majoris, Vmag = −1.46), to the dimmest, Regulus (Alpha Leo, Vmag = 1.35, Table 2). Many single stars used for calendric purposes in Aboriginal astronomical traditions are first magnitude[4] (Clarke, 2007; Hamacher and Norris, 2011; Johnson, 1998). Celestial objects that are dimmer than this magnitude limit tend to be star clusters (e.g. Pleiades), the Milky Way (including the dark spaces within), close pairs of stars (e.g. Shaula and Lesath in Scorpius), or stars that form an asterism (e.g. Crux, belt and scabbard of Orion). Therefore, we assume Kaurna seasonal stars are first magnitude.

Table 2: First magnitude stars, their Bayer designation, their apparent brightness (Vmag). The Vmag for each star is their average apparent magnitude in the absence of an atmosphere (not their apparent magnitude when at an altitude of 5°), taken from SIMBAD catalogue (simbad.u-strasbg.fr/simbad/).

| Name | Bayer Designation | $V_{mag}$ |
|---|---|---|
| Sirius | Alpha Canis Majoris | −1.46 |
| Canopus | Alpha Carinae | −0.72 |
| Rigil Kent | Alpha Centauri | −0.29 |
| Arcturus | Alpha Boötis | −0.04 |
| Vega | Alpha Lyrae | 0.03 |
| Capella | Alpha Auriga | 0.08 |
| Rigel | Beta Orionis | 0.12 |
| Procyon | Alpha Canis Minoris | 0.34 |
| Achernar | Alpha Eridani | 0.46 |
| Betelgeuse | Alpha Orionis | 0.45 |
| Hadar | Beta Centauri | 0.61 |
| Altair | Alpha Aquilae | 0.77 |
| Acrux | Alpha Crucis | 0.77 |
| Aldebaran | Alpha Taurii | 0.85 |
| Spica | Alpha Virginis | 0.98 |
| Antares | Alpha Scorpii | 1.06 |
| Pollux | Beta Geminorum | 1.14 |
| Fomalhaut | Alpha Piscis Austrini | 1.16 |
| Mimosa | Beta Crucis | 1.25 |
| Deneb | Alpha Cygni | 1.25 |
| Regulus | Alpha Leonis | 1.36 |

### 4.1.2 Stellar and Solar Position

The minimum altitude of a star when it is first visible to the naked eye depends on the brightness of the star, the colour of the star, the elevation of the observer, the atmospheric conditions between the star and the observer, and the presence and phase of the moon. Under good seeing conditions, a human observer with very good visual acuity can detect stars as faint as magnitude 8 (Kelley and Milone, 2011; Schaefer, 2000)

Assuming the observer is at sea-level, looking to an unobstructed horizon in very good seeing conditions, a first magnitude star needs to have a minimum altitude of 5° to be visible (*ibid*: 125). Otherwise, extinction of the star's light is too great. The sun needs to have a maximum altitude of −10° for a first magnitude star to be visible when it rises (Aveni, 2001: 112). If the sun's altitude is greater than this limit, its light will drown out the light of the star.

Heliacal and acronycal rising only applies to objects with an azimuth between 0° (due north) and 180° (due south). The appearance of stars nearer to the sun will be more greatly reduced by the sun's light than stars with an azimuth further from the sun, such as those rising acronycally.

Stars that are circumpolar as seen from Adelaide ($\delta \leq -54.5°$) are always visible above the horizon and therefore do not "rise."

### 4.2    Selection Criteria

Seasonal change occurs on a timescale of weeks to months. Therefore, minor factors affecting the calculation of heliacal and acronycal rise times by a few days can be ignored for this study. The interested reader is referred to Purrington (1988) and Schaefer (2000) for detailed analyses of heliacal rising calculations.

Using the major factors for calculating heliacal and acronycal rise times, calendric stars must meet the following four criteria:

1. The star must be first magnitude.
2. The star must not be circumpolar as seen from Adelaide ($\delta \leq -54.5°$)
3. The star must have an azimuth between 0° and 180° at the time of heliacal or acronycal rising.
4. The star must have a minimum altitude of 5° when the sun has a maximum altitude of −10°.

Because early ethnographers sometimes conflated "star", "planet", and "constellation", we are unsure if some seasonal stars are in fact a star or an asterism/constellation. We consider the latter case, provided that at least one star in the asterism is first magnitude. It should be noted that connect-the-dots constellations in Aboriginal astronomical traditions are rare. Clarke (2014: 318) suggests that Aboriginal observers focused on individual elements, such as stars, rather than drawing lines between them to make shapes.

### 4.3    Estimating Seasonal Markers

The Kaurna traditions do not claim seasonal change occurs exactly when the star is first visible. In the case of Parna, it serves as a warning of the coming wet season, allowing the people a "grace period" during which they need to prepare.

The association between stars and seasons are less clear with Wolta and Wilto, and non-existent for Kudlilla. We use the description given in Kaurna traditions (if any) and combine that with temperature and rainfall data from the region to approximate the time of year we expect to see the calendar star rise.

When estimating the time of month we search for heliacally rising seasonal stars, the following definitions are used:

- *Early* = 1st to 10th of the month
- *Mid* = 11th to 20th of the month.
- *Late* = 21st to 30th/31st of the month.

### 4.4 Comparative Studies

Another line of evidence for identifying celestial objects in Kaurna traditions is to compare the roles of these stars and their terrestrial counterparts in nearby Aboriginal astronomical traditions. Although distinct Aboriginal languages and cultures are distinct, many have similar linguistic themes and related astronomical associations, particularly in southeastern Australia (see Clarke, 2007; Fredrick, 2008; Hamacher, 2012; Johnson, 1998; Tindale, 1983). [5]

## 5 Kaurna Seasonal Stars

### 5.1 Parna – The Autumn Star

In the Kaurna language recorded by Teichelmann and Schürmann (1840: 37), *parna* has three different meanings. It can signify (1) the personal pronoun "they", (2) one of the two men placed at either side of a line the people form when about to perform a circumcision as part of a male ceremony (the other man, Tappo, is a fly), or (3) the "autumn star", which is derived from *Parnatti*, meaning "autumn, when the star Parna is seen."

Autumn on the Adelaide Plain is denoted by increased rainfall, which peaks during the winter months. Building waterproof homes (colloquially called "wurlies", Figure 5) was essential for the Kaurna to survive the cold, wet winter months (Taplin, 1879: 126). This is due, in part, to water run-off from the Mount Lofty Ranges to the east, flooding the relatively flat plains that drain into Gulf St Vincent to the west (Figure 1: Right).

The most significant rise in rainfall in terms of the percentage difference between any particular month and the previous month occurs in March (Table 3). Rain increases by an average of 20 mm each month until it peaks in June. The records indicate that the appearance of the star Parna provides enough time for the Kaurna to build their huts before the heavy rains arrive. This suggests that Parna appears at the start of March, in agreement with Heyes (1999: 17). Teichelmann and Schürmann (1840: 50) say people begin building huts during *Wadlaworngatti* - "the beginning of April (autumn season)." Heyes suggests that the confusion may be due to Teichelmann and Schürmann conflating Parnatti and Wadlaworngatti into a single "autumn" season stretching from late February to early June.

Teichelmann and Schürmann do not clarify the time of day the star Parna is seen; only that it signals the start of autumn.

In the nearby Yaraldi dialect of the Ngarrindjeri language, the term for autumn is *marangalkadi*, which means "autumn, time of the crow" (raven) – from February to April - and *parnar* means 'rain'. The raven is a prominent figure in Ngarrindjeri

Dreamings (Berndt et al., 1993: 21, 76, 240-242). In Ngarrindjeri culture, the "autumn stars" are low in the southeastern sky because the crow-spirit entered the skyworld to the southeast of the Lower Murray, towards Mount Gambier in the far southeast corner of South Australia (Clarke, 1997: 137). Although the Ngarrindjeri language is very different to the Kaurna language, there are cultural relationships between the groups. It is unknown if this plays a significant part in the identity of Parna, but the link is plausible.

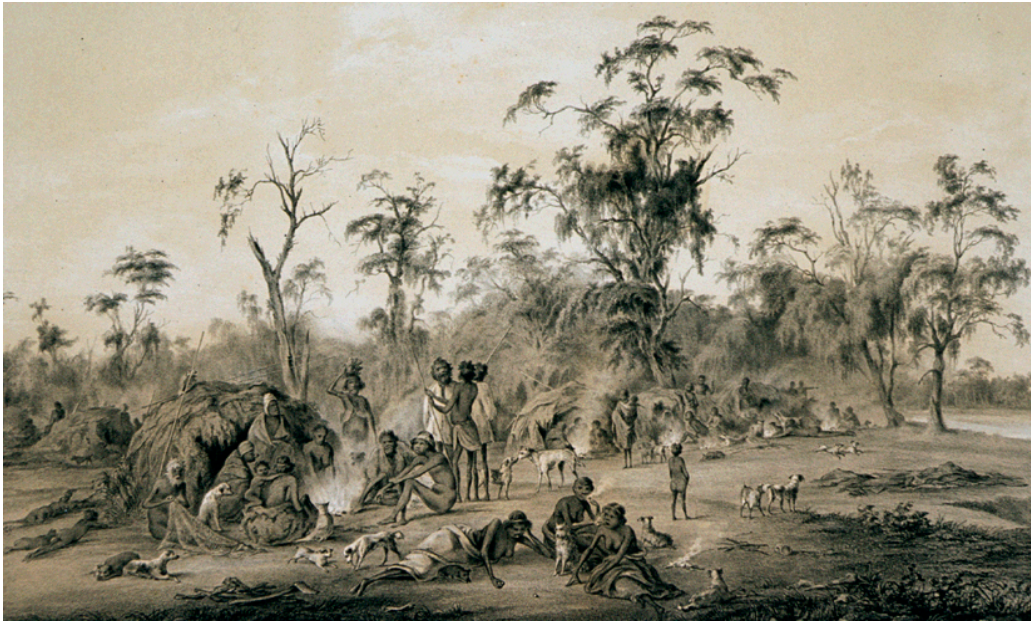

Figure 5: Aboriginal huts, or "wurlies" constructed by the Kaurna as shelter during the rainy winter season. Image: Alexander Schramm, South Australian Museum.

Table 3: The mean and median rainfall (in millimeters) in the Adelaide Plain, taken from the data in Figure 3. The percentage change in rainfall between one month and the previous month are given.

| Month | Mean Rainfall | Percentage Change | Median Rainfall | Percentage Change |
|---|---|---|---|---|
| January | 18.9 | 32.7% | 20.1 | 15.5% |
| February | 15.8 | 16.4% | 6.8 | 66.2% |
| March | 27.0 | −70.9% | 19.0 | −179.4% |
| April | 40.1 | −48.5% | 33.6 | −76.8% |
| May | 60.7 | −51.4% | 56.8 | −69.0% |
| June | 80.6 | −32.8% | 76.8 | −35.2% |
| July | 77.5 | 3.8% | 68.8 | 10.4% |
| August | 69.0 | 11.0% | 69.2 | −0.6% |
| September | 58.4 | 15.4% | 58.0 | 16.2% |
| October | 42.2 | 27.7% | 37.0 | 36.2% |
| November | 30.3 | 28.2% | 30.3 | 18.1% |
| December | 28.1 | 7.3% | 23.8 | 21.5% |

Given the information above, we search for stars that meet the criteria in both early March and early April. Ngarrindjeri traditions indicate that Parna may have a southeasterly azimuth (between 90° and 180°). Heyes (1999: 17) identifies early-March as the time Parna appears in the sky.

### 5.1.1 Place names indicating Parna

Place names incorporate Parna and its stellar significance. *Parnangga*, a hilltop campsite at Morphett Vale, in the southern suburbs of Adelaide, allegedly means "place of the autumn star." The suffix *–ngga* is locative (Amery, 2002: 167). Parnangga may refer to the appearance of Parna and/or 'place of the procession leader' (Schultz, 2013a). The latter probably relates to male circumcision, a common component of manhood rites and male initiation. In the nearby Nukunu culture, 'partnapa' was the 'first stage of initiation; young initiate who has gone through this' (Hercus, 1992).

Amery (2002) suggests that Parnangga may be a 'stepping-off' place (typically mountains or hills) where an ancestor ascended into the sky. The campsite is believed to be near the primary school at Morphett Vale East, although this is uncertain (Schultz, 2013a). Considering the topography of the area, such as the elevation of Parnangga and the maximum elevation of the nearby Lofty Ranges, a star with a minimum altitude of 5° will be visible above the ranges as seen from Parnangga. Tindale (c.1931) and Cooper (1949: 19, 20) relate Parna to Parnangga, which they claim means 'autumn rains' or 'place of autumn rains.' This is probably based on the transcription error of Meyer (1843: 90), who confused the Ngarrindjeri word *Parnar* (rain) for the Kaurna word for rain (*kuntoro*; Teichelmann and Schürmann, 1840: 14). The Ngarrindjeri are south of Adelaide, along the Coorong, and speak a very different language to Kaurna.

According to Tindale (c.1931), a stream near Second Valley on the Fleurieu Peninsula south of Adelaide named *Parananacooka* translates to "excreta and urine of the Autumn Star women, so called because of the intense brackishness of the river at the end of summer". Schultz (2013b) and Clarke (1997: 145) both claim Tindale's derivation is speculative. The interested reader is directed to Schultz (2013a,b) for a detailed etymology of the place names Parnangga and Parananacooka.

### 5.2   Kudlilla – The Winter Star?

Kudlilla is "winter", the rainy season (Teichelmann and Schürmann, 1840: 12). It is not associated with a star in the recorded literature. For this study, we propose it *could* be and predict it will rise heliacally at the beginning of winter. The mean monthly rainfall is highest from June to August, which corresponds to the three coldest months of the year (in terms of mean daily temperature).

For this reason, we search for stars that meet the criteria in late May or early June.

### 5.3   Wilto – The Spring Star

Teichelmann and Schürmann (1840: 55) record *Wilto* as a species of eagle and a star, and *Wiltutti* as a season (its corresponding time of year is not provided), *Willutti* as the "spring" season, and *Willo* as a star (and also refers to one whose elder brother has died). It is probable the latter two terms are contractions of the former (MacPherson, 1881: 79), meaning the season is equivalent to spring and is associated with the eagle (most likely the wedge-tailed eagle, *Aquila audax*).

As with Wolta, little information is provided in the literature about the appearance of either star or its relationship to either season (or the eagle). In the Turra (Narangga) traditions of Yorke Peninsula (bordering Kaurna country to the west), Wiltu is the eaglehawk (wedge-tailed eagle) (Fison and Howitt, 1880: 285; Tindale, 1936: 67).

Although Teichelmann and Schürmann (1840: 55) refer to Wilto as a star (singular), Clarke (1990: 137) suggests that Wilto refers to the brightest stars of the Southern Cross (Crux), as it shares linguistic similarities with Ngadjuri traditions in north-central South Australia where the eagle, Wildu, is associated with Crux (regarded as the footprint of the eagle). Crux is also related to the wedge-tailed eagle in other Aboriginal traditions, such as the Adnyamathanha of South Australia (Hamacher et al., 2015; Johnson, 1998). Crux is at its highest altitude (crossing the meridian) at dawn in mid-January and its lowest altitude at dawn in early July. Crux is circumpolar as seen from Kaurna country (the minimum altitude of Gamma Crucis in 1840 as seen from Adelaide is ~ 1.5°), and was thus rejected from the criteria.

Across Aboriginal Australia, stars are connected to animals because the rise/set times of stars at dusk or dawn correlate to some aspect of that animal's behaviour. This might include the animals' migration patterns, nest building, breeding, whelping, and brooding, among other aspects. Therefore, we examine eagle behaviour to see if any aspects correspond to the heliacal rising of first magnitude stars. This is conjecture, but may provide some insight as to why the star is associated with a particular animal. The study of star rise/set times and their animal counterparts in Indigenous traditions is the focus of ongoing research by Leaman and Hamacher (2015) and Leaman et al. (2015).

The breeding cycle of the wedge-tailed eagle begins in March and April and runs to September, but a majority of the eagles lay their eggs in July (Olsen, 1995, 2005). Mid-July is the middle of winter, thus making it a poor "spring star". Eagles incubate their eggs for 45 days and brood the chicks for an additional 30 days. Therefore, most young eagles are leaving their nests in mid-September (11$^{th}$-20$^{th}$), which is also the end of the greater breeding season and the middle of spring. Heyes (1999: 17) cites the appearance of the Wilto star in early to mid-September, though he does not associate this with the animal's behaviour.

Given this information, we search for a star that meets the criteria in mid-September, keeping in mind that linguistic evidence supports Crux as Wilto.

**5.4    Wolta – The Summer Star**

Wolta, the "wild turkey" star, is associated with "summer" (Worltatti). Little information is provided about the appearance of Wolta or its relationship to summer. Summer is the hottest and driest season of the year, with the highest temperatures occurring in January and February, with nearly equal mean temperatures in December and March.

The terms "brush turkey", "plains turkey", and "wild turkey" are commonly used by Aboriginal people to describe the Australian bustard (*Ardeotis australis*), which is found across the Adelaide Plains. Walta/Waltja is also the name of the bustard in the nearby Narranga language of southern Yorke Peninsula (Tindale, 1936: 67). In the Nukunu language, *Waalha* refers to the 'plains turkey' (Hercus, 1992: 29) and in the Adnyamathanha language of the Flinders Ranges north of Adelaide, *Waltha* refers to the 'monsoonal summer rain' (Schebeck, 1974).

The relationship between the star and the animal are unknown, but probably relate to some aspect of the bird's behaviour, such as the breeding cycle or seasonal migrations. The Australian bustard breeds once a year. In South Australia, eggs are laid from July to November (Boehm, 1947), but tend to occur primarily from September/October to November/December, and the egg incubates for 23-24 days (Beruldsen, 2003). The bustard is diurnal, being active in the early morning and late evenings. During the start of the breeding season, the males begin a powerful courtship display.

Detailed study of bustard behaviour in Victoria (Ziembicki, 2009) shows that the bustard returns to breeding ground in spring and summer as more food is available (following the wet winter rains). Anecdotal observations suggest that numbers peak during the summer in Victoria, implying possible movements southwards from inland regions at this time (*ibid:* 28). The bustard also prefers to feed on grasshoppers in the summer in Victoria (*ibid:* 188). No similar, in-depth studies have been conducted in Adelaide, so any comparison with bustard behaviour in Victoria is conjecture.

This poses a challenge to clearly identify the time of year this star might rise heliacally or acronycally. For this study, we search for stars that rise in early to mid-December, coinciding with the end of the bustard breeding season and the start of the hot, dry summer. In agreement, Heyes (1999: 17) cites early to mid-December as the appearance of Wolta.

## 6 Results

All first magnitude stars that rise heliacally and acronycally during the year as seen from Adelaide are given in Table 4. Stars that meet the criteria for each seasonal star for both heliacal and acronycal rising are given. Those in green are the best candidates, and those in red are within two days of the time period explored for each candidate (i.e. early, mid, late month).

### 6.1 Parna

Three stars are possible contenders for Parna: Fomalhaut and Deneb, depending on whether it coincides with early March or early-April (Heyes, 1999 suggests the former: Figure 4), and Vega, as it is within two days of early March.

The stars Fomalhaut and Vega appear in the sky at close to the same time, with Vega appearing in the northeast and Fomalhaut in the southeast. Deneb appears in early-April in the far northeastern sky. Because of the declination of Fomalhaut (−29.61°), it also heliacally sets on the same days it heliacally rises from the latitude of Adelaide. This means Fomalhaut has an altitude of 5° when the sun has an altitude of −10° at both dusk and dawn around the first of March. This is not the case with either Vega or Deneb.

In the nearby Mallee Scrub of southeast South Australia and western Victoria, Vega is associated with the mallee fowl (*Leipoa ocellata*), a ground dwelling megapode (mound-building bird) in Wergaia traditions (Stanbridge, 1861). The acronycal rising of Vega coincides with the time of year the birds are building their nesting mounds. The star's heliacal setting coincides with the time the chicks begin hatching from the eggs.

Table 4: Stars visible from Adelaide that meet the criteria for dates in 1840 (when Teichelmann and Schürmann published their work on the Adelaide Aboriginal language). The dates given are the first day the criteria are met (using Stellarium: www.stellarium.org), The azimuth is rounded up to the nearest degree. Stars in green meet the criteria. Stars in red "marginally" meet the criteria, meaning they are within two days of the estimated start or end period.

| Star | Az (°) | Heliacal Rise | | Acronycal Rise | |
|---|---|---|---|---|---|
| | | Date | Candidate | Date | Candidate |
| Arcturus | 61 | 17-Dec | Wolta | 18-May | Kudlilla |
| Altair | 76 | 16-Feb | | 29-Jul | |
| Vega | 35 | 27-Feb | Parna | 11-Aug | |
| Fomahlaut | 124 | 1-Mar | Parna | 14-Aug | |
| Deneb | 23 | 4-Apr | Parna | 18-Sep | Wilto |
| Canopus | 156 | 10-May | | 22-Oct | |
| Rigel | 97 | 5-Jun | Kudlilla | 15-Nov | |
| Aldebaran | 66 | 12-Jun | Kudlilla | 21-Nov | |
| Sirius | 106 | 21-Jun | | 28-Nov | Wolta |
| Betelgeuse | 77 | 24-Jun | | 2-Dec | Wolta |
| Procyon | 79 | 21-Jul | | 22-Dec | |
| Capella | 20 | 7-Aug | | 4-Jan | |
| Pollux | 50 | 17-Aug | | 11-Jan | |
| Regulus | 71 | 18-Sep | Wilto | 8-Feb | |
| Spica | 99 | 8-Nov | | 30-Mar | Parna |
| Antares | 119 | 16-Dec | Wolta | 17-May | |

Fomalhaut meets the description of an autumn star in Yaraldi (Ngarrindjeri) traditions, as it appears in the southeastern sky (both Deneb and Vega are northerly stars with an Az < 35°). Fomalhaut also rises and sets heliacally on the same days, unlike any of the other calendric stars describes in Table 4.

With this information, Fomalhaut is the most plausible candidate for Parna. Tindale (1934) originally identified Parna as the Pleiades, but the heliacal rising of that cluster occurs in June and the Pleiades are also already identified in Kaurna astronomical traditions as the Mankamankarranna.

### 6.2  Kudlilla

In terms of heliacal rise, only one candidate star meets the criteria for Kudlilla (Rigel), although Aldebaran is near the boundary of the upper-limit of the time period for Kudlilla. In terms of acronycal rise, Arcturus marginally qualifies. When comparing candidate stars with those identified in Kaurna traditions, we see that Rigel is already associated with Parnakkoyerli. Scott Heyes does not include Kudlilla as a seasonal star in his analysis of the Kaurna calendar (Figure 4).

Both Kudlilla and Parna have marginal associations with Arcturus and Spica, respectively.

### 6.3    Wilto

Regulus meets the heliacal rising criteria and Deneb meets the acronycal rising criteria. For reasons described in Section 4.3, Wilto is probably the Southern Cross (which did not meet the criteria set out for this study, as none of the stars in Crux are first magnitude).

### 6.4    Wolta

Two stars meet the criteria for heliacal rise in mid-December: Antares and Arcturus. Betelgeuse meets the criteria for acronycal rise, and Sirius marginally does. It should be noted that Arcturus, Antares, and Betelgeuse are all visibly red stars and this may somehow relate to the hot, dry season as red stars commonly relate to fire.

## 7    Discussion and Concluding Remarks

The identification of only one seasonal Kaurna star is supported by this study with multiple lines of evidence: Parna. Parna is identified as Fomalhaut as it meets the criteria and is at such a declination that it heliacally rises *and* sets at the same time of year – the start of the autumn season.

Additional circumstantial evidence is found by examining Ngarrindjeri traditions to the south. In Ngarrindjeri traditions, the autumn stars (Marangani/Marangalkadi, Western counterpart unidentified) are visible in the southeastern sky towards Mount Gambier. Autumn is the time of the crow (Australian Raven) and lasts from February to April (Berndt *et al.*, 1993: 76). Records of both the Kaurna and Ngarrindjeri indicate that they have four main seasons, each denoted by a star or group of stars (*ibid*). This suggests that if these two groups possess some similarities in their astronomical traditions, Parna may also appear in the southeastern sky.

It should be emphasized that the Kaurna and Ngarrindjeri languages are very different, so traditions from both groups cannot be easily conflated and any connection is speculative without further supporting evidence. A more detailed study of Ngarrindjeri astronomical traditions is the focus of future work.

Fomalhaut is found in the astronomical traditions of other Aboriginal cultures of Australia, though rarely (Clarke, 2007: 48; Hamacher and Frew, 2010: 223; Johnson, 1998):

- In Wotjobaluk (Wergaia) and Mara (Gunditjmara) traditions of western Victoria, Fomalhaut is the moiety eaglehawk (wedge-tailed eagle) ancestor (Massola, 1968: 109), despite not being included in the detailed Boorong (Wergaia) study by Stanbridge (1861).

- In Wurundjeri traditions of central Victoria, Fomalhaut is Bunjil, the primary sky-hero and "all father" (Dawson, 1881: 100; Hartland, 1898: 306; Howitt, 1884a,b: 452; Howitt, 1904: 489), although Bunjil is also related to the star Altair (*ibid*).

- In Bundjalung traditions of northern coastal NSW, Fomalhaut is *Bunnungar*, the "frill-necked lizard" (Patston, 1997), a probable reference to the Eastern bearded dragon (*Pogona barbata*).

- Similarly, in Euahlayi traditions of north-central NSW, Fomalhaut is *Gani*, a "small iguana" (Brough-Smyth, 1878: 286; Ridley, 1875: 142). This is probably a reference to a small species of goanna.

- In Wardaman traditions of the north-central Northern Territory, Fomalhaut is *Menggen*, the white cockatoo, whose feathers were used for ceremonial decoration (Cairns and Harney, 2003: 115, 126, 138, 204). Menggen watches over ceremonies and is part of a complex songline. Senior Wardaman custodian, Bill Yidumduma Harney, is from the Menngen (White Cockatoo) community.

- In the Torres Strait, Fomalhaut was called *Panauna graz* (Rivers, 1912: 224), but no further details about its meaning are provided[7].

A peculiar trend in the records of Aboriginal astronomical traditions across Australia is that several first magnitude stars are generally absent. For example, the first magnitude stars omitted by Stanbridge (1861) in his study of Boorong (Wergaia) traditions of western Victoria are Procyon, Betelgeuse, Spica, Fomalhaut, Deneb, and Regulus. Curiously, these are all potential seasonal stars (Table 4).

MacPherson (1881: 73-75) provides a possible explanation for the omission of some of these first magnitude stars in Aboriginal traditions: Fomalhaut, Spica, and Regulus (as well as Procyon) are "isolated" stars that do not fit into any "mechanical" grouping, an apparently common occurrence in the Aboriginal astronomical traditions of nearby western Victoria.

In this grouping, MacPherson proposes that characters represented by stars of a particular family are either: (A) Grouped based upon their arrangement in the sky, specifically grouping three stars (or clusters) in a linear pattern; (B) grouped into four linear arrangements that are roughly parallel to each other; or (C) arranged roughly parallel to the horizon as they rise in the evening sky in their respective seasons at the latitude of the region (36° S) (after Hamacher and Frew, 2010).

Perhaps one reason these stars do not appear in Aboriginal traditions is that certain stars are considered secret and sacred. For example, an Aboriginal man was once pressed as to why Procyon was unnamed in his people's astronomical traditions. When asked, he "merely shook his head in a negative sort of way" (EKV, 1884).

Considering the diversity of Aboriginal cultures (> 350 distinct languages) and the length of time they have been in Australia (> 50,000 years), relatively little research has been conducted on Aboriginal astronomical traditions, and much of that was recorded by amateur (and professional) ethnographers with limited astronomical training. It is possible that the omission of these stars reflects the incompleteness of recorded Aboriginal astronomical traditions. Further research will shed light on the issue.

Using a combination of linguistic analysis, heliacal and acronycal rise calculations, seasonal weather data, and comparative studies of Aboriginal astronomical traditions, we attempt to identify Kaurna seasonal stars.

Only one star was supported by more than one line of evidence: Fomalhaut, the autumn-star Parna. The star is uncommon in recorded Aboriginal astronomical traditions, which poses important considerations for future work.

## 9      Acknowledgements

The author acknowledges and pays respect to Kaurna elders past and present and would like to thank Rob Amery, Mark Carter, Philip Clarke, Paul Curnow, Gail Higginbottom, Trevor Leaman, Stephen McCluskey, Ray Norris, Wayne Orchiston, and Chester Schultz for feedback, comments, and criticisms.

This research made use of the Stellarium (stellarium.org), SIMBAD stellar database (Strasbourg), Astrophysics Database System (Harvard), Bureau of Meteorology (Australia), South Australian Museum (Adelaide), Library of South Australia (Adelaide), and the TROVE database (Canberra).

The author acknowledges support from Australian Research Council grant DE140101600.

## 10     Notes

1. The term Kaurna was not used at the time of colonisation to refer to the Aboriginal people of the Adelaide Plains. It was later popularized by Norman Tindale in the 1920s, resulting in its adaptation by Adelaide Plains traditional owners and its common usage today (Amery, 2000; Clarke, 1997). A comprehensive database of information on the Kaurna language can be found here: http://www.mobilelanguageteam.com.au/languages/kaurna

2. Sometimes ethnographers conflate the terms "star" and "constellation", such as "Wayakka" a Kaurna word for an unidentified star or constellation (Teichelmann and Schürmann, 1840: 55).

3. Weather data (1977-2014) from Adelaide. Taken from www.weatherzone.com.au/climate/station.jsp?lt=site&lc=23090

4. There are exceptions to this, as one would expect. But this tends to be the case overall.

5. Kaurna astronomical traditions are not well known to non-initiated Kaurna men, only a small portion of which have ever been recorded (Curnow, 2006: 8).

6. In Aboriginal astronomical traditions across Australia, the stars of Orion's belt and scabbard commonly represent a group of boys/men and the Pleiades commonly represent a group of girls/women (see Clarke, 2007; Fuller, et al. 2014; Hamacher, 2012; Johnson, 1998; Leaman and Hamacher, 2014; Tindale, 1983). A detailed study of the Pleiades and Seven Sisters Dreamings in Aboriginal traditions is the focus of doctoral candidate Melissa Razuki at RMIT in Melbourne.

7. *Panauna* was not defined or otherwise mentioned in any Torres Strait Islander records, but *graz* is a fishtrap or weir built of stones on a reef (Ray, 1907: 99). This term was recorded on Mabuiag Island, suggesting it is from the Kalaw Lagaw Ya language.

## 11    References

Amery, R.M., 2000. *Warrabarna Kaurna! - Reclaiming an Australian Language*. Lisse, The Netherlands, Swets & Zeitlinger.

Amery, R.M. 2002. *Weeding out Spurious Etymologies: Toponyms on the Adelaide Plains*. In *The land is a map: placenames of Indigenous origin in Australia*, edited by L. Hercus, F. Hodges, and J. Simpson. Canberra, Pandanus Books in association with Pacific Linguistics, Australian National University, pp. 165-180.

Aveni, A.F. 2001. *Sky Watchers of Ancient Mexico, 2nd Edition*. Austin, University of Texas Press.

Aveni, A.F., 2003. Archaeoastronomy in the Ancient Americas. *Journal of Archaeological Research*, 11(2), 149-191.

Berndt, R.M., Berndt, C.H., and Stanton, J.E., 1993. *A World that Was: The Yaraldi of the Murray River and the Lakes, South Australia.* Melbourne, Melbourne University Press.

Beruldsen, G.R., 2003. *Australian Birds: Their Nests and Eggs*. Brisbane, Gordon R. Beruldsen.

Black, J.M., 1920. Vocabularies of four South Australian languages. *Transactions of the Royal Society of South Australia*, 44, 76-93

Boehm, E.F., 1947. The Australian bustard: with special reference to its past and present status in south Australia. *South Australian Ornithologist*, 18, 37–40.

Brough-Smyth, R., 1878. *The Aborigines of Victoria, Vol. II*. London, John Ferres.

Cairns, H.C., and Harney, B.Y., 2003. *Dark Sparklers*. Merimbula, NSW, Hugh C. Cairns.

Clarke, P.A., 1990. Adelaide Aboriginal Cosmology. *Journal of the Anthropological Society of South Australia*, 28, 1-10.

Clarke, P.A., 1997. The Aboriginal Cosmic Landscape of Southern South Australia. *Records of the South Australian Museum*, 29(2), 125-145

Clarke, P.A., 2007. An Overview of Australian Aboriginal Ethnoastronomy. *Archaeoastronomy*, 21, 39-58.

Clarke, P.A., 2014. The Aboriginal Australian cosmic landscape. Part I: the ethnobotany of the Skyworld. *Journal of Astronomical History and Heritage*, 17(3), 307–335.

Cooper, H.M., 1949. *Australian Aboriginal words and their meanings*. Adelaide, South Australian Museum.

Curnow, P., 2006. Kaurna Night Skies. *Bulletin of the Astronomical Society of South Australia*, 115(6), 8-10.

Dawson, J., 1881. *Australian Aborigines - The Language and Customs of several tribes of Aborigines in the Western District of Victoria, Australia*. Sydney, George Robertson.

E.K.V., 1884. Aboriginal Star Knowledge. *The Queenslander*, Saturday, 6 September 1884, p. 387.

Fison, L., and Howitt, A.W., 1880. *Kamilaroi and Kurnai*. Melbourne, George Robertson.

Fredrick, S., 2008. *The Sky of Knowledge - A Study of the Ethnoastronomy of the Aboriginal People of Australia*. Master of Philosophy Thesis, School of Archaeology & Ancient History, University of Leicester, Leicester, UK.

Fuller, R.S., Norris, R.P., and Trudgett, M., 2014. The Astronomy of the Kamilaroi People and their Neighbours. *Australian Aboriginal Studies*, 2014(2), 3-27.

Gell, J.P., 1842. The vocabulary of the Adelaide tribe. *Tasmanian Journal of Natural Science & Agricultural Statistics*, 1, 109-124.

Hamacher, D.W., 2012. *On the Astronomical Knowledge and Traditions of Aboriginal Australians*. Doctor of Philosophy Thesis, Department of Indigenous Studies, Macquarie University, Sydney, Australia.

Hamacher, D.W. and Frew, D.J., 2010. An Aboriginal record of the Great Eruption of Eta Carinae. *Journal of Astronomical History and Heritage*, 13(3), 220-234.

Hamacher, D.W., Leaman, T.M., and Carter, M., 2015. Aboriginal Astronomical traditions from Ooldea, South Australia, Part II: Animals in the Ooldean Sky. *Journal of Astronomical History and Heritage*, in preparation.

Hamacher, D.W., and Norris, R.P., 2011. *Bridging the Gap through Australian Aboriginal Astronomy*. In "*Archaeoastronomy & Ethnoastronomy - Building Bridges Between Cultures*", edited by C.L.N. Ruggles. Cambridge, Cambridge University Press, pp. 282-290.

Hartland, E., 1898. The "High Gods" of Australia. *Folklore*, 9(4), 290-329.

Hercus, L.A., 1992. *A Nukunu Dictionary*. Canberra, Australian Institute for Aboriginal and Torres Strait Islander Studies.

Heyes, S., 1999. *The Kaurna Calendar: Seasons of the Adelaide Plains*. Honours Paper, School of Architecture, Landscape Architecture, and Urban Design, University of Adelaide, South Australia.

Howitt, A.W., 1884a. On Some Australian Beliefs. *Journal of the Anthropological Institute*, 13, 185-198.

Howitt, A.W., 1884b. On Some Australian Ceremonies of Initiation. *The Journal of the Anthropological Institute of Great Britain and Ireland* 13: 432–459.

Howitt, A.W., 1904. *The Native Tribes of Southeast Australia*. London, MacMillan.

Johnson, D.D., 1998. *The Night Skies of Aboriginal Australia - a Noctuary*. Sydney, University of Sydney Press.

Kelley, D. and Milone, E., 2011. *Exploring Ancient Skies - a survey of ancient and cultural astronomy, 2nd edition*. New York, Springer.

Leaman, T.M., and Hamacher, D.W., 2014. Aboriginal Astronomical traditions from Ooldea, South Australia, Part I: Nyeeruna and the Orion Story. *Journal of Astronomical History and Heritage*, 17(2), 180-194.

Leaman, T.M., Hamacher, D.W., and Carter, M. 2015. Aboriginal Astronomical traditions from Ooldea, South Australia, Part II: Animals in the Ooldean sky. *Journal of Astronomical History and Heritage*, in preparation.

Leaman, T.M., and Hamacher, D.W., 2015. Methodologies for correlating animal behaviour to stellar positions in Australian Aboriginal astronomies. *In preparation.*

MacPherson, P., 1881. Astronomy of the Australian Aborigines. *Journal and Proceedings of the Royal Society of New South Wales,* 15, 71-80.

Manning, G.H., 2002. [*Aboriginal Australians 1837-1858: An Essay - Aborigines on the Coastal Plain*](). Adelaide, State Library of South Australia.

Massola, A., 1968. *Bunjil's Cave*. Melbourne, Lansdowne.

Meyer, H.A.E., 1843. *Vocabulary of the Language Spoken by the Aborigines of South Australia.* Adelaide, Allen.

Olsen, P., 1995. *Australian Birds of Prey: the Biology and Conservation of Raptors*. Sydney, University of New South Wales Press.

Olsen, P., 2005. *Wedge-Tailed Eagle*. Australian Natural History Series. Melbourne, CSIRO Publishing.

Patston, G., 1997. *Dreamtime Stories for Bedtime Reading*. Woodbridge (Tasmania),

Dover.

Purrington, R.D., 1988. Heliacal rising and setting: quantitative aspects. *Journal for the History of Astronomy* (*Archaeoastronomy Supplement)*, 12, S72-S84.

Ray, S., 1907. Cambridge Anthropological Expedition to Torres Straits Vol. III – Linguistics. Cambridge, Cambridge University Press.

Ridley, W., 1875. *Kamilaroi and Other Australian Languages*. Sydney, Thomas Richards.

Rivers, W.H.R., 1912. *Reports of the Cambridge Anthropological Expedition to Torres Straits. Vol. IV - arts and crafts*. Cambridge, Cambridge University Press.

Schaefer, B.E., 2000. New Methods and Techniques for Historical Astronomy and Archaeoastronomy. *Archaeoastronomy*, 15, 121-136.

Schebeck, B., 1974. *Texts on the Social System of the Atynyamatana People with Grammatical Notes*. Canberra, Pacific Linguistics, Series D 21, Research School of Pacific Studies, Australian National University.

Schultz, C., 2013a. *Place Name Summary (PNS) 4.1.2/04: Parnanngga*. Kaurna Warra Pintyandi - The Southern Kaurna Place Names Project, University of Adelaide.

Schultz, C., 2013b. *Place Name Summary (PNS) 5.1.2/02: Parananacooka*. Kaurna Warra Pintyandi - The Southern Kaurna Place Names Project, University of Adelaide.

Schürmann, C.W., 1840. The Aborigines of South Australia. *The South Australian Colonist*, Tuesday, 10 March 1840.

Stanbridge, W.E., 1861. Some Particulars of the General Characteristics, Astronomy, and Mythology of the Tribes in the Central Part of Victoria, Southern Australia. *Transactions of the Ethnological Society of London*, 1, 286-304.

Taplin, G., 1879. *The Folklore, Manners, Customs, and Languages of the South Australian Aborigines*. Adelaide, E. Spiller.

Teichelmann, C.G., 1841. The Aborigines of South Australia. *Southern Australian* (newspaper published in Adelaide from 1838 to 1844), Tuesday, 20 April 1841, p. 4

Teichelmann, C.G. 1857. *Dictionary of the Adelaide dialect*. No. 59, Bleek's Catalogue of Sir George Grey's Library dealing with Australian languages, South African Public Library.

Teichelmann, C.G., and Schürmann, C.W., 1840. *Outlines of a grammar, vocabulary, and phraseology, of the Aboriginal language of South Australia.* Adelaide, Teichelmann and Schürmann.

Tindale, N.B. c.1931-c.1991. *Place Names: N.B. Tindale Ms SE of S Australia*. AA 338/7/1/44. Adelaide, South Australian Museum Archives.

Tindale, N.B., 1934. *Tindale annotated map, Hundred of Yankalilla.* AA 338/24/101. Adelaide, South Australian Museum.

Tindale, N.B., 1936. Notes on the Natives of the Southern Portion of Yorke Peninsula, South Australia. *Transactions and proceedings of the Royal Society of South Australia*, 60, 55-70.

Tindale, N.B., 1937. Two Legends of the Ngadjuri Tribe from the Middle North of South Australia. *Transactions of the Royal Society of South Australia*, 61, 149-153.

Tindale, N.B., 1974. *Aboriginal Tribes of Australia.* Berkeley, University of California Press.

Tindale, N.B., 1983. Celestial lore of some Australian Aboriginal tribes. *Archaeoastronomy*, 12/13, 258-379.

Ziembicki, M., 2009. *Ecology and movements of the Australian bustard Ardeotis australis in a dynamic landscape.* Doctor of Philosophy Thesis, School of Earth and Environmental Sciences, University of Adelaide, Adelaide, South Australia.

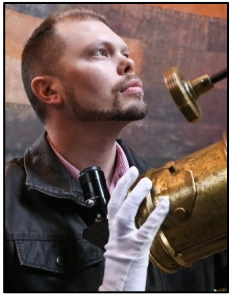

Dr Duane Hamacher is a Lecturer and ARC Discovery Early Career Research Fellow at the University of New South Wales in Sydney, Australia. His research and teaching focuses on cultural astronomy. He leads the Indigenous Astronomy group at UNSW, is a council member of the International Society of Archaeoastronomy and Astronomy in Culture (ISAAC), and is an Associate Editor of the Journal of Astronomical History and Heritage.